\documentclass[10pt]{article}

\usepackage[T1]{fontenc}
\usepackage[utf8]{inputenc}
\usepackage{lmodern}
\usepackage[margin=0.85in]{geometry}
\usepackage{amsmath,amssymb,mathtools}
\usepackage{graphicx}
\usepackage{float}
\usepackage{booktabs}
\usepackage{array}
\usepackage{microtype}
\usepackage{xcolor}
\usepackage{cite}
\usepackage[colorlinks=true,linkcolor=blue!55!black,citecolor=blue!55!black,
            urlcolor=blue!55!black]{hyperref}
\hypersetup{
  pdftitle={Cross-Moment Consistency and the Fifth Adler Coefficient in Hadronic Tau Decays},
  pdfauthor={Gauhar Abbas, Debabina Biswas, and Vartika Singh}
}

\graphicspath{{figures/}{./}}

\newcommand{\MSbar}{\overline{\mathrm{MS}}}
\newcommand{\Dhat}{\widehat D}
\newcommand{\chat}{\widehat c}
\newcommand{\dhat}{\widehat d}

\newcommand{\Wtau}{W_\tau}
\newcommand{\WGC}{W_{\rm GC}}
\newcommand{\WP}{W_{P2}}
\newcommand{\cO}{\mathcal O}
\newcommand{\dd}{\mathrm d}

\newcommand{\arxiv}[1]{\href{https://arxiv.org/abs/#1}{arXiv:#1}}
\newcommand{\doi}[1]{\href{https://doi.org/#1}{doi:#1}}

\title{\vspace{-2.0em}\bfseries
Cross-Moment Consistency and the Fifth Adler Coefficient in Hadronic Tau Decays}

\author{Gauhar Abbas\footnote{gauhar.phy@itbhu.ac.in},
Debabina Biswas\footnote{debabinabiswas.rs.phy24@itbhu.ac.in}, Pratham Dadhich\footnote{pratham.rs.phy24@itbhu.ac.in }, and
Vartika Singh\footnote{vartikasingh.rs.phy19@itbhu.ac.in}\\[3pt]
\normalsize\itshape Department of Physics, Indian Institute of Technology
(BHU), Varanasi 221005, India}
\date{November 4, 2026}

\begin{document}
\maketitle

\begin{abstract}
The unknown coefficient $c_{5,1}$ controls the first omitted fifth-order term
in fixed-order analyses of inclusive hadronic $\tau$ decays.  We determine
what the four known Adler coefficients can actually test.  For any normalized
polynomial weight, the fixed-order Adler-to-moment map is triangular with unit
diagonal.  Hence one common Adler-coefficient error reaches every moment with
the same absolute size, and a multi-weight relative test reduces to one
absolute tolerance.  For three representative weights, a ten-percent test
resolves $c_{5,1}$ only to about $\pm24$; the closest contour-improved test is
weaker by more than an order of magnitude.  Moreover, any continuation that
preserves the observed sign of the direct Pad\'e/Shanks defects gives
$c_{5,1}\geq378.03$, so values near $280$ require additional structure.  A
calibration-and-falsification protocol applied to a simplified
Beneke--Jamin-type pure-power ansatz gives the conditional estimate
$c_{5,1}=248\pm30\,(\mathrm{struct})\pm74\,(\mathrm{method})=248\pm80$.
At the physical order its central value coincides algebraically with the exact
four-parameter fit; the protocol adds a falsifiable admissibility test and a
method uncertainty, not new coefficient information.  We also
clarify that the often-quoted Beneke--Jamin value $283$ is a linear
extrapolation of total-width FOPT term ratios, not an output of their main
five-parameter renormalon model.
\end{abstract}

% \noindent\textbf{Keywords:} hadronic $\tau$ decays; Adler function;
% spectral moments; Pad\'e approximation; renormalons; operator product expansion

\section{Introduction}
\label{sec:intro}

Inclusive hadronic $\tau$ decay is a low-energy precision laboratory for
QCD~\cite{Davier:2005xq,Pich:2020gzz}.  The inclusive width and weighted
integrals of the measured invariant-mass spectrum can be related, by
analyticity, to contour integrals of current
correlators~\cite{Narison:1988ni,Braaten:1991qm}.  On most of the contour the
relevant scale is of order $m_\tau$, so perturbative QCD can be combined with
the operator product expansion (OPE)~\cite{Wilson:1969zs} to extract the
strong coupling.  The perturbative contribution is known through
$\cO(\alpha_s^4)$ for massless three-flavor QCD~\cite{Baikov:2008jh}, and
this calculation has underpinned a series of determinations of
$\alpha_s(m_\tau)$~\cite{Davier:2008sk,Maltman:2008nf,Beneke:2008ad,
Bethke:2009jm,Pich:2010xb,Pich:2011cj,Bethke:2011tr,Boito:2011qt,Boito:2014sta,
Pich:2016bdg,Boito:2020xli,Ayala:2022cxo,Boito:2025pwg}.  Power corrections
in the OPE are numerically suppressed for the total
width~\cite{Braaten:1991qm,Davier:2005xq,Davier:2008sk,Beneke:2008ad,
Pich:2010xb}, whereas the treatment of quark--hadron duality violations near
the timelike axis remains an active
issue~\cite{Cata:2008ye,Boito:2011qt,Boito:2014sta,Pich:2016bdg,Pich:2022tca,
Boito:2024gtb}.  The next Adler coefficient, conventionally denoted $c_{5,1}$,
has not been computed directly.

The practical significance of $c_{5,1}$ is limited but real.  It controls the
first omitted coefficient in a fifth-order fixed-order expansion.  It does
\emph{not}, by itself, decide the all-order difference between fixed-order
perturbation theory (FOPT) and contour-improved perturbation theory
(CIPT)~\cite{Pivovarov:1991rh,LeDiberder:1992jjr}, which has long been the dominant
theoretical uncertainty in $\alpha_s$
extractions~\cite{Jamin:2005ip,Beneke:2008ad,Menke:2009vg,Cvetic:2010ut,
Groote:2012jq,Beneke:2012vb}.  That difference originates in the asymptotic
nature of the perturbative series, whose large-order behaviour is governed by
renormalons~\cite{Beneke:1998ui}; it depends on the running of the coupling
along the complex contour and on the singularity structure of the Borel
transform~\cite{Hoang:2020mkw,Hoang:2021nlz,Golterman:2023oml,Gracia:2023qdy}.
Subtracting the gluon-condensate renormalon reduces it
substantially in favor of FOPT~\cite{Benitez-Rathgeb:2022yqb,Benitez-Rathgeb:2022hfj,
Beneke:2023wkq,Beneke:2025hlg}.

In the normalization used throughout this article, the reduced massless Adler
function at a Euclidean scale $Q^2$ is
\begin{equation}
 \Dhat(Q^2)=\sum_{n=1}^{\infty}c_{n,1}\,a(Q^2)^n,
 \qquad a(Q^2)\equiv\frac{\alpha_s(Q^2)}{\pi},
 \label{eq:adler-series}
\end{equation}
and the four known coefficients for $n_f=3$ in the $\MSbar$
scheme~\cite{Bardeen:1978yd} are~\cite{Baikov:2008jh}
\begin{equation}
 c_{1,1}=1,\qquad c_{2,1}=1.63982,\qquad
 c_{3,1}=6.37101,\qquad c_{4,1}=49.0757.
 \label{eq:known-coefficients}
\end{equation}
The size of the unknown higher orders~\cite{Pich:2020gzz,David:2013gaa} has
been estimated with scheme-invariant (effective-charge and
minimal-sensitivity) methods~\cite{Kataev:1995vh},
explicit renormalon models of the Adler
function~\cite{Beneke:2008ad,Caprini:2009vf,DescotesGenon:2010cr,
Jamin:2010cv,Caprini:2010tf,Caprini:2011ya,Beneke:2012vb},
Pad\'e and Borel--Pad\'e approximants~\cite{Boito:2018rwt}, conformal
mappings of the Borel plane~\cite{Caprini:2019kwp}, and nonlinear sequence
transformations~\cite{Abbas:2025dpm}.  More broadly, the perturbative series has been reorganized by
renormalon-chain resummation~\cite{LovettTurner:1994hx,Ball:1995ni,
LovettTurner:1995ti,Neubert:1995gd,Maxwell:1996he,Maxwell:1997yw,
Maxwell:2001he}, modified Borel and ultraviolet-renormalon
studies~\cite{Altarelli:1994vz,Cvetic:2001ws}, renormalization-group-summed and
conformally mapped expansions~\cite{Abbas:2012py,Abbas:2012fi,Abbas:2013usa,
Caprini:2020lff,Caprini:2023tfa,Caprini:2024unl}, series
acceleration~\cite{Schilcher:2024fjx}, and scheme or scale
variations~\cite{Boito:2016pwf,Wu:2018cmb,Caprini:2018agy,Wu:2019mky}.

Producing a number for $c_{5,1}$ is therefore not the difficulty; many methods
do so.  The question addressed here is what the four known coefficients can
actually \emph{test}.  Our conclusions are:
\begin{enumerate}
 \item Testing a common Adler forecast on several normalized FOPT moments is
 exactly one scalar test, because the Adler-to-moment map has unit diagonal
 (Sections~\ref{sec:transfer} and~\ref{sec:tolerance}).
 \item At order five, the commonly used ten-percent, three-weight test permits
 an absolute error of about $24$ in $c_{5,1}$; the corresponding CIPT test is
 weaker by more than an order of magnitude (Section~\ref{sec:cipt}).
 \item A direct geometric Shanks/Pad\'e continuation gives $c_{5,1}=378.03$,
 and every continuation of the Adler sequence that preserves the sign of the
 observed forecast defects gives $c_{5,1}\geq378.03$, above the usual
 phenomenological range (Sections~\ref{sec:pade} and~\ref{sec:obstruction}).
 \item A value near $280$ therefore requires additional structure that reverses
 the defect sign.  A simplified Beneke--Jamin (BJ)-type renormalon
 ansatz~\cite{Beneke:2008ad} can supply it, provided the distinction between
 calibration and independent validation is retained together with its model
 dependence (Sections~\ref{sec:renormalons}
 and~\ref{sec:cvrc}).
\end{enumerate}
Sections~\ref{sec:setup} and~\ref{sec:weights} set up the moments and explain
the choice of weights.  Section~\ref{sec:long-paper-guide} explains what the
companion long paper~\cite{CompanionPaper} adds beyond this self-contained
account.

\section{From the measured spectrum to an Adler moment}
\label{sec:setup}

\subsection{What the experiment measures}

For a vector or axial-vector quark current $J_\mu$, we define
\begin{equation}
 \Pi_{\mu\nu}(q)=i\int \dd^4x\,e^{iq\cdot x}
 \langle0|T\{J_\mu(x)J_\nu(0)^\dagger\}|0\rangle .
 \label{eq:correlator}
\end{equation}
Lorentz covariance separates transverse ($J=1$) and longitudinal ($J=0$)
scalar correlators~\cite{Braaten:1991qm}.  Their discontinuities define spectral functions,
\begin{equation}
 \rho^{(J)}(s)=\frac{1}{\pi}\operatorname{Im}\Pi^{(J)}(s+i0).
 \label{eq:spectral-function}
\end{equation}
Experimentally, exclusive hadronic final states are reconstructed, detector
efficiencies and backgrounds are corrected, and migration between true and
reconstructed invariant-mass bins is unfolded.  The useful data product is
therefore a vector of spectral-bin contents together with its full covariance
matrix, not only a plotted curve~\cite{ALEPH:2005qgp,Davier:2013sfa,
Boito:2020xli,Boito:2025pwg}.  A weighted moment
can be formed as
\begin{equation}
 I_{\rm exp}^{(w)}(s_0)=\frac{1}{s_0}\int_0^{s_0}\dd s\,
 w\!\left(\frac{s}{s_0}\right)\rho_{\rm exp}(s),
 \qquad s_0\leq m_\tau^2,
 \label{eq:experimental-moment}
\end{equation}
or, for bins $i$ of width $\Delta s_i$,
\begin{equation}
 I_w\simeq\sum_i q_i^{(w)}\rho_i,
 \qquad q_i^{(w)}=\frac{\Delta s_i}{s_0}w(x_i).
 \label{eq:binned-moment}
\end{equation}
If $C_{ij}=\operatorname{Cov}(\rho_i,\rho_j)$, linear error propagation gives
\begin{equation}
 \operatorname{Cov}(I_w,I_v)=
 \sum_{i,j}q_i^{(w)}C_{ij}q_j^{(v)}.
 \label{eq:moment-covariance}
\end{equation}
We notice that different moments are generally strongly correlated because they use the same
events, the same unfolding response, and common normalization and systematic
corrections.  Thus many polynomial weights do not automatically provide the
same number of independent experimental measurements.

The physical transverse kinematic weight in the total hadronic $\tau$ width is
\begin{equation}
 w_\tau(x)=(1-x)^2(1+2x)=1-3x^2+2x^3,
 \qquad x=\frac{s}{m_\tau^2}.
 \label{eq:physical-spectral-weight}
\end{equation}
It is fixed by the spin structure and phase space of the
decay~\cite{Braaten:1991qm}, not selected to optimize perturbation theory.  Other weights define additional moments of
the differential spectrum; they are not alternative expressions for the
unmodified total width.

\subsection{Finite-energy sum rule and integration by parts}

Analyticity of $\Pi(s)$ away from its physical cut implies the finite-energy
sum rule (FESR)~\cite{Braaten:1991qm,Pich:2020gzz}
\begin{equation}
 \frac{1}{s_0}\int_0^{s_0}\dd s\,w(s/s_0)\rho(s)
 =-\frac{1}{2\pi i s_0}\oint_{|s|=s_0}\dd s\,
 w(s/s_0)\Pi(s),
 \label{eq:fesr}
\end{equation}
with a counterclockwise contour.  Polynomial weights are convenient because
they are analytic within the contour and project OPE terms transparently.  An
integration by parts replaces the correlator by the Adler function.  If
$x=s/s_0$ and the spectral weight is $w(x)$, we define the corresponding Adler
weight
\begin{equation}
 W(x)=2\int_x^1\dd z\,w(z),
 \label{eq:weight-relation}
\end{equation}
and normalize the weights so that $W(0)=2\int_0^1\dd z\,w(z)=1$.  Applying
Eq.~\eqref{eq:weight-relation} to Eq.~\eqref{eq:physical-spectral-weight}
gives
\begin{equation}
 \begin{split}
 W_\tau(x)
 &=2\int_x^1\dd z\,(1-3z^2+2z^3)\\
 &=1-2x+2x^3-x^4=(1-x)^3(1+x).
 \end{split}
 \label{eq:Wtau-derivation}
\end{equation}
The distinction matters: $w_\tau$ weights the measured spectrum, whereas
$W_\tau$ multiplies the Adler function on the complex contour.

The perturbative Adler moment is
\begin{equation}
 \delta_W^{(0)}(s_0)=\frac{1}{2\pi i}
 \oint_{|x|=1}\frac{\dd x}{x}\,W(x)\Dhat(-s_0x).
 \label{eq:adler-moment}
\end{equation}
For the total nonstrange $V+A$ width, this quantity enters schematically as
\begin{equation}
 R_{\tau,V+A}=3|V_{ud}|^2S_{\rm EW}
 \left[1+\delta_{\rm EW}'+\delta^{(0)}
 +\sum_{D\geq2}\delta^{(D)}\right],
 \label{eq:Rtau-structure}
\end{equation}
where $S_{\rm EW}$ and $\delta_{\rm EW}'$ are electroweak
corrections~\cite{Marciano:1988vm,Braaten:1990ef,Davier:2002dy},
$\delta^{(0)}\equiv\delta^{(0)}_{\Wtau}(m_\tau^2)$ is the massless
perturbative term, and $\delta^{(D)}$ denotes mass and power corrections of
OPE dimension $D$~\cite{Braaten:1991qm,Pich:1999hc}.

\section{Why these three weights are complementary}
\label{sec:weights}

We use the normalized Adler weights
\begin{align}
 \Wtau(x)&=(1-x)^3(1+x)=1-2x+2x^3-x^4,
 \label{eq:Wtau}\\
 \WGC(x)&=(1-x)^3=1-3x+3x^2-x^3,
 \label{eq:WGC}\\
 \WP(x)&=(1-x)^2(1+x)=1-x-x^2+x^3.
 \label{eq:WP2}
\end{align}
The label ``GC'' indicates gluon-condensate sensitivity.  The label ``$P2$''
is a mnemonic for a twice-pinched polynomial: $\WP$ has a double zero at
$x=1$.  Endpoint zeros (``pinching'') suppress the region near the timelike point
$s=s_0$, where the OPE is least reliable and duality violations can be most
visible~\cite{Cata:2008ye,Boito:2011qt,Pich:2022tca,Boito:2024gtb}.  More
pinching can therefore reduce one source of uncertainty, but a higher-degree
polynomial also projects higher-dimensional OPE terms.  Weight design is a
trade-off, not a rule that more zeros are always better.

The three weights also probe the leading infrared renormalon
differently~\cite{Beneke:2008ad,Beneke:2012vb}.  If
$W(x)=\sum_{m=0}^M W_mx^m$, the Borel transform of the moment contains the
kinematic kernel
\begin{equation}
 K_W(u)=\frac{\sin(\pi u)}{\pi}
 \sum_{m=0}^{M}\frac{W_m}{u-m}.
 \label{eq:borel-kernel}
\end{equation}
To evaluate it at a positive integer $p$, we set $u=p+\varepsilon$.  Then
\begin{equation}
 \frac{\sin(\pi u)}{\pi}=(-1)^p\varepsilon+\cO(\varepsilon^3),
 \qquad
 \sum_m\frac{W_m}{u-m}=\frac{W_p}{\varepsilon}
 +\sum_{m\ne p}\frac{W_m}{p-m}+\cO(\varepsilon).
 \label{eq:kernel-expansion}
\end{equation}
Multiplication gives the simple selection rule
\begin{equation}
 K_W(p)=(-1)^pW_p.
 \label{eq:kernel-selection}
\end{equation}
If $W_p=0$, the kernel instead has a zero proportional to $u-p$.  Therefore a
simple Adler Borel pole at $u=p$ is cancelled, while a branch point is weakened
by one power.

At the gluon-condensate location $u=2$ we have,
\begin{equation}
 (W_\tau)_2=0,\qquad (W_{\rm GC})_2=3,\qquad (W_{P2})_2=-1.
 \label{eq:u2-coefficients}
\end{equation}
The standard total-$\tau$ weight suppresses the leading $u=2$ infrared
renormalon, associated with the gluon condensate~\cite{Beneke:1998ui}, and is
therefore an unusually forgiving extrapolation
target~\cite{Beneke:2008ad,Beneke:2012vb}.
$W_{\rm GC}$ and $W_{P2}$ retain the same singularity with different signs and
normalizations.  This is the physical reason to compare them: the set contains
one renormalon-suppressed moment and two renormalon-sensitive moments.  The
comparison is informative about cancellations and relative sensitivity, even
though, as the next two sections show, it does not turn a common coefficient
forecast into three independent constraints.

\section{The triangular FOPT map and the transfer identity}
\label{sec:transfer}

In FOPT the running coupling on the contour is re-expanded in powers of
$a_0=a(s_0)$.  Renormalization-group invariance fixes every logarithmic term
in terms of lower-order Adler coefficients and the beta
function~\cite{Beneke:2008ad,Baikov:2016tgj}, so that
\begin{equation}
 \Dhat(-s_0x)=\sum_{N=1}^{\infty}a_0^N
 \sum_{n=1}^{N}c_{n,1}P_{Nn}\!\left(\ln(-x)\right),
 \qquad P_{NN}=1.
 \label{eq:RG-reexpansion}
\end{equation}
Two properties of this expansion are used below.  It is triangular,
$n\leq N$, because $a^n(-s_0x)$ starts at $a_0^n$; and $P_{NN}=1$, because
that leading term carries no contour logarithm.  Inserting
Eq.~\eqref{eq:RG-reexpansion} in Eq.~\eqref{eq:adler-moment} and integrating
term by term with the logarithmic contour moments
\begin{equation}
 J_k[W]=\frac{1}{2\pi i}\oint_{|x|=1}\frac{\dd x}{x}\,W(x)\ln^k(-x)
 \label{eq:log-moments}
\end{equation}
gives the FOPT series
\begin{equation}
 \delta_W^{\rm FOPT}(s_0)=\sum_{N=1}^{\infty}d_N[W]\,a_0^N,
 \qquad
 d_N[W]=\sum_{n=1}^{N}M_{Nn}[W]\,c_{n,1}.
 \label{eq:triangular-map}
\end{equation}
The off-diagonal entries $M_{Nn}$, $n<N$, are combinations of beta-function
coefficients and $J_{k\geq1}[W]$ that can be evaluated in closed
form~\cite{Beneke:2008ad,Penarrocha:2001ig}.  The diagonal follows from
$P_{NN}=1$ and Cauchy's formula,
\begin{equation}
 M_{NN}[W]=J_0[W]=\frac{1}{2\pi i}\oint\frac{\dd x}{x}\,W(x)=W(0)=1,
 \label{eq:unit-diagonal}
\end{equation}
since every weight in Eqs.~\eqref{eq:Wtau}--\eqref{eq:WP2} is normalized to
$W(0)=1$.  Separating the diagonal term,
\begin{equation}
 d_N[W]=c_{N,1}+g_N[W],\qquad
 g_N[W]=\sum_{n=1}^{N-1}M_{Nn}[W]\,c_{n,1},
 \label{eq:gap}
\end{equation}
where the RG-generated part $g_N[W]$ is fixed entirely by the lower Adler
coefficients.

Let a hat denote a forecast.  Keep $c_{1,1},\ldots,c_{N-1,1}$ at their exact
values and forecast only $c_{N,1}$.  Transporting this common forecast through
Eq.~\eqref{eq:triangular-map} gives $\dhat_N[W]=\chat_{N,1}+g_N[W]$, and
subtracting Eq.~\eqref{eq:gap},
\begin{equation}
 \dhat_N[W]-d_N[W]=M_{NN}[W]\,(\chat_{N,1}-c_{N,1})
 =\chat_{N,1}-c_{N,1}
 \qquad\text{for every }W.
 \label{eq:transfer-identity}
\end{equation}
This is the transfer identity.  It is exact under its hypothesis: one common
Adler forecast is transported through the same FOPT map with all lower inputs
unchanged.  It is not a relation between forecasts obtained independently
from each moment series, and it concerns \emph{absolute} errors.  Relative
errors differ because the denominators differ; at $N=4$,
\begin{equation}
 d_4[\Wtau]=127.079,\qquad d_4[\WGC]=214.140,\qquad d_4[\WP]=21.509,
 \label{eq:d4-values}
\end{equation}
so the small $\WP$ coefficient magnifies any common discrepancy.

\subsection{Provenance of the Beneke--Jamin values 52 and 283}
\label{sec:bj-values}

Beneke and Jamin (BJ)~\cite{Beneke:2008ad} observed that the convergence rate
of the FOPT total-width series, with terms $u_n=d_n[\Wtau]\,a_0^n$ at their
input $a_0=0.34/\pi$, is nearly uniform.  The
term ratios
$r_n=u_n/u_{n-1}=a_0\,d_n[\Wtau]/d_{n-1}[\Wtau]$ are
\begin{equation}
 (r_2,r_3,r_4)=(0.5630,\;0.5485,\;0.5216),
\end{equation}
each close to one half, with a slight order dependence that BJ fitted
linearly.  Their estimate is therefore a linear extrapolation of the term
ratios, not a constant-ratio (geometric) continuation.  With
$r_2$ and $r_3$ the line passes through two points, $\widehat r_4=2r_3-r_2
=0.5340$.  With $r_2$, $r_3$, $r_4$ the least-squares line has slope
$(r_4-r_2)/2$ and passes through the mean at $n=3$, so
$\widehat r_5=(4r_4+r_3-2r_2)/3=0.5030$, consistent with the value quoted in
Ref.~\cite{Beneke:2008ad}.  Since every $r_n$ is proportional to $a_0$, so
is the fitted line, and $a_0$ cancels in
$\dhat_{N+1}=\widehat r_{N+1}d_N/a_0$.
Writing $d_n\equiv d_n[\Wtau]$,
\begin{equation}
 \dhat_4=d_3\left(\frac{2d_3}{d_2}-\frac{d_2}{d_1}\right)=130.09,
 \qquad
 \dhat_5=\frac{d_4}{3}\left(\frac{4d_4}{d_3}+\frac{d_3}{d_2}
 -\frac{2d_2}{d_1}\right)=590.6 ,
 \label{eq:bj-closed-form}
\end{equation}
with $d_2/d_1=5.2023$, $d_3/d_2=5.0681$ and $d_4/d_3=4.8198$.  Subtracting
the RG-generated parts $g_4[\Wtau]=78.003$ and $g_5[\Wtau]=307.783$
[Eqs.~\eqref{eq:gap} and \eqref{eq:d5-affine}] gives
\begin{equation}
 \chat_{4,1}=130.0867-78.0029=52.0838\simeq52,
 \qquad
 \chat_{5,1}=590.6056-307.7825=282.8231\simeq283,
 \label{eq:bj-c5}
\end{equation}
the postdiction of footnote~2 and the estimate of Eq.~(3.10) of
Ref.~\cite{Beneke:2008ad}.  Both numbers extrapolate the FOPT total-width
coefficients in moment space and are converted to Adler coefficients only
afterwards; neither is a direct continuation of the common Adler sequence.
In particular, $283$ is an input to the main five-parameter BJ renormalon
model, not an output of that fit.  When BJ instead drop this input, set the
linear polynomial term to zero, and fit a four-parameter model to
$c_{1,1},\ldots,c_{4,1}$, that separate construction predicts about $280$
(our reconstruction gives $280.12$).

The postdiction illustrates the transfer identity.  Using the unrounded
reconstruction $\chat_{4,1}=52.0838$ and the exact $c_{4,1}=49.0757$ gives the
same absolute error $3.0081$ in every moment, by
Eq.~\eqref{eq:transfer-identity}, but the relative errors are
\begin{equation}
 \frac{3.0081}{127.079}=2.367\%,\qquad
 \frac{3.0081}{214.140}=1.405\%,\qquad
 \frac{3.0081}{21.509}=13.986\%
 \label{eq:numerical-transfer}
\end{equation}
for $\Wtau$, $\WGC$ and $\WP$.  A statement such as ``the forecast works for
$\Wtau$ but fails for $\WP$'' therefore describes the relative-error
convention, not three different underlying coefficient errors.  Judged by the
three-weight criterion of Section~\ref{sec:tolerance}, the estimator that
produces $283$ fails at the order where it can be checked:
$|\chat_{4,1}-c_{4,1}|=3.008$ exceeds the tolerance $2.151$ by a factor
$1.399$.  Using the published rounded postdiction $52$ would instead give
$2.9243$ and $1.36$; mixing those rounded quantities with the unrounded
formula is the source of the discrepancy in the earlier version.

\section{Resolving power of a multi-weight test}
\label{sec:tolerance}

Let $\mathcal W$ be a set of weights normalized to $W(0)=1$ and impose a
relative tolerance $\epsilon$ on every moment,
\begin{equation}
 |\dhat_N[W]-d_N[W]|\leq\epsilon\,|d_N[W]|,
 \qquad W\in\mathcal W.
 \label{eq:relative-criterion}
\end{equation}
By the transfer identity~\eqref{eq:transfer-identity} all left-hand sides are
equal, so the intersection of these conditions is exactly
\begin{equation}
 |\chat_{N,1}-c_{N,1}|\leq
 \epsilon\min_{W\in\mathcal W}|d_N[W]|.
 \label{eq:one-tolerance}
\end{equation}
A multi-weight FOPT test is therefore a single absolute tolerance on one
number.  An additional weight tightens it only if its moment coefficient is
smaller in magnitude; it cannot probe a new direction, because the
forecast-error space is one-dimensional.  The collapse relies on
holding the lower inputs fixed: if several Adler coefficients or
beta-function inputs vary together, the error vector
$\delta d_N[W]=\sum_n M_{Nn}[W]\,\delta c_{n,1}$ can have rank larger than
one, and sufficiently different weights then do provide independent
constraints.  Likewise, the experimental covariance of the moments,
Eq.~\eqref{eq:moment-covariance}, measures how much independent information
the data contain, which is a different question from how one missing
perturbative coefficient propagates; a phenomenological fit should retain
both.

Table~\ref{tab:fopt-tolerance} lists the coefficients for the three weights of
Section~\ref{sec:weights}.  At every order the smallest one, and hence the
tolerance, belongs to $\WP$.  By the unit diagonal, the order-five
coefficients are exactly affine in the unknown $c_{5,1}$,
\begin{equation}
 d_5[\Wtau]=c_{5,1}+307.783,\qquad
 d_5[\WGC]=c_{5,1}+1156.925,\qquad
 d_5[\WP]=c_{5,1}-512.597,
 \label{eq:d5-affine}
\end{equation}
so the fifth-order row is shown at the representative value $c_{5,1}=277$,
the central value of the Borel--Pad\'e estimate of Ref.~\cite{Boito:2018rwt};
another choice shifts all three entries rigidly.

\begin{table}[t]
\centering
\caption{FOPT moment coefficients and the coefficient tolerance
$\epsilon\min_W|d_N[W]|$ of Eq.~\eqref{eq:one-tolerance} at $\epsilon=0.1$.
The $N=5$ row uses $c_{5,1}=277$.  The minimum is set by $\WP$ at every
order.}
\label{tab:fopt-tolerance}
\begin{tabular}{ccccc}
\toprule
$N$ & $d_N[\Wtau]$ & $d_N[\WGC]$ & $d_N[\WP]$ & coefficient tolerance\\
\midrule
$3$ & $26.366$ & $34.484$ & $14.523$ & $\pm1.45$\\
$4$ & $127.079$ & $214.140$ & $21.509$ & $\pm2.15$\\
$5$ & $584.78$ & $1433.93$ & $-235.60$ & $\pm23.6$\\
\bottomrule
\end{tabular}
\end{table}

$\WP$ remains the most restrictive weight at fifth order whenever
$|c_{5,1}-512.597|<|c_{5,1}+307.783|$, i.e.\ for $c_{5,1}>102.4$, which covers
the phenomenological range.  There Eq.~\eqref{eq:one-tolerance} reads
\begin{equation}
 |\chat_{5,1}-c_{5,1}|\leq0.1\,|c_{5,1}-512.597|
 \simeq24\quad(c_{5,1}=277),
 \label{eq:fifth-order-resolution}
\end{equation}
ranging from $18$ to $29$ over the Borel--Pad\'e range $c_{5,1}=277\pm51$.  The tolerance shrinks to
zero as $c_{5,1}$ approaches $512.6$, where $d_5[\WP]$ vanishes and a relative
criterion for that moment ceases to be meaningful.  Two candidate values of
$c_{5,1}$ closer than this tolerance cannot be distinguished, however many
weights display the same common error.  This is a statement about the
validation procedure, not a claim that the moments carry no distinct physics.

To set the scale of the fifth-order term, we take
$\alpha_s(M_Z)=0.11873$~\cite{Brida:2025gii}, as in Ref.~\cite{Abbas:2025dpm},
and evolve it with RunDec~\cite{Chetyrkin:2000yt,Herren:2017osy} at five loops
with $\MSbar$ decoupling of the bottom and charm quarks at their
$\MSbar$ masses, obtaining $\alpha_s(m_\tau)=0.3196$, so that $a=0.10173$ and
$a^5=1.090\times10^{-5}$.  This input is used only here and in
Section~\ref{sec:cipt}; the reconstruction of Ref.~\cite{Beneke:2008ad} in
Section~\ref{sec:bj-values} uses that paper's own value $a_0=0.34/\pi$.  Because
$M_{55}=1$, a change $\Delta c_{5,1}$ shifts $d_5[W]$ by the same amount for
every weight.  For $\Delta c_{5,1}=80$, the total uncertainty of the forecast
in Eq.~\eqref{eq:cvrc-result}, the fifth-order contribution to each FOPT
moment changes by
\begin{equation}
 \Delta c_{5,1}\,a^5\simeq8.7\times10^{-4},
 \label{eq:numerical-impact}
\end{equation}
about $0.4\%$ of $\delta^{\rm FOPT}_{\Wtau}\simeq0.203$ (evaluated through
fifth order at $c_{5,1}=277$).  The coefficient
$c_{5,1}$ thus matters for precision theory, but its uncertainty is only one
component of the complete experimental and theoretical error budget.

\section{What changes in contour-improved perturbation theory}
\label{sec:cipt}

FOPT re-expands the running coupling in $a(s_0)$ before performing the
contour integral.  CIPT~\cite{Pivovarov:1991rh,LeDiberder:1992jjr} instead
keeps the running coupling on the contour and truncates the Adler series
before integration:
\begin{equation}
 \delta^{\rm CIPT}_{W,N}=\sum_{n=1}^{N}c_{n,1}J_n^{\rm CI}[W],
 \qquad
 J_n^{\rm CI}[W]=\frac{1}{2\pi i}\oint\frac{\dd x}{x}\,W(x)\,a^n(-s_0x).
 \label{eq:cipt-definition}
\end{equation}
This map is diagonal in the Adler coefficients, but its entries are
$\alpha_s$-dependent contour integrals rather than one.  A forecast error at
the last retained order therefore enters as
\begin{equation}
 \widehat\delta^{\rm CIPT}_{W,N}-\delta^{\rm CIPT}_{W,N}
 =(\chat_{N,1}-c_{N,1})\,J_N^{\rm CI}[W],
 \label{eq:cipt-error}
\end{equation}
and the simultaneous relative test again collapses to a single inequality,
\begin{equation}
 |\chat_{N,1}-c_{N,1}|\leq
 \epsilon\min_{W\in\mathcal W}
 \left|\frac{\delta^{\rm CIPT}_{W,N}}{J_N^{\rm CI}[W]}\right| .
 \label{eq:cipt-tolerance}
\end{equation}
The unit diagonal of Eq.~\eqref{eq:unit-diagonal} is lost, but the logical
collapse survives: one unknown coefficient multiplies one known factor for
each weight.

Table~\ref{tab:fopt-cipt} compares the two tests weight by weight.  The FOPT
entries are $\epsilon|d_N[W]|$ from Table~\ref{tab:fopt-tolerance}; they
involve only pure numbers and do not depend on $\alpha_s$.  The CIPT entries
are the right-hand side of Eq.~\eqref{eq:cipt-tolerance} for each weight
separately, obtained by integrating the five-loop renormalization-group
equation on the contour at $\alpha_s(m_\tau)=0.3196$.  The $N=5$ entries use
$c_{5,1}=277$ in both schemes.  The tolerance of the three-weight test is the
smallest entry in each row (in bold).

\begin{table}[ht]
\centering
\caption{Single-weight coefficient tolerances at $\epsilon=0.1$.  FOPT:
$\epsilon|d_N[W]|$.  CIPT: $\epsilon|\delta^{\rm CIPT}_{W,N}/J^{\rm CI}_N[W]|$
at $\alpha_s(m_\tau)=0.3196$.  The $N=5$ entries use $c_{5,1}=277$.  Bold
entries are the three-weight tolerances.}
\label{tab:fopt-cipt}
\begin{tabular}{c ccc ccc}
\toprule
 & \multicolumn{3}{c}{FOPT} & \multicolumn{3}{c}{CIPT}\\
\cmidrule(lr){2-4}\cmidrule(lr){5-7}
$N$ & $\Wtau$ & $\WGC$ & $\WP$ & $\Wtau$ & $\WGC$ & $\WP$\\
\midrule
4 & $\pm12.7$ & $\pm21.4$ & $\boldsymbol{\pm2.15}$
  & $\pm123$ & $\boldsymbol{\pm77.7}$ & $\pm227$\\
5 & $\pm58.5$ & $\pm143$ & $\boldsymbol{\pm23.6}$
  & $\pm1600$ & $\boldsymbol{\pm770}$ & $\pm5465$\\
\bottomrule
\end{tabular}
\end{table}

The most restrictive weight differs between the schemes, for different
reasons.  In FOPT it is the weight with the smallest $|d_N[W]|$, which is $\WP$
because its RG-generated part is negative and partly cancels the Adler
coefficient: $d_4[\WP]=c_{4,1}-27.567$ and $d_5[\WP]=c_{5,1}-512.597$.  In
CIPT the partial sums differ only moderately between weights,
$\delta^{\rm CIPT}_{W,4}=0.182$, $0.223$, $0.144$ for $\Wtau$, $\WGC$, $\WP$,
whereas the last-order contour integrals differ much more,
$J^{\rm CI}_4[W]=(1.48,\,2.87,\,0.63)\times10^{-4}$ and
$J^{\rm CI}_5[W]=(1.16,\,3.01,\,0.26)\times10^{-5}$.  The tightest CIPT bound
therefore comes from $\WGC$, which has the largest $J^{\rm CI}_N$, and $\WP$,
the most restrictive weight in FOPT, becomes the least restrictive in CIPT.
The comparison does not rely on mixing weights: for $\WGC$ alone, FOPT is
tighter than CIPT by factors of $3.6$ at $N=4$ and $5.4$ at $N=5$.

CIPT is less selective because the last retained term is only a small fraction
of the contour-improved partial sum.  For $\WGC$ it is $6.3\%$ of
$\delta^{\rm CIPT}_{W,4}$ and $3.6\%$ of $\delta^{\rm CIPT}_{W,5}$, so a
ten-percent tolerance on the partial sum admits an excursion larger than the
coefficient being tested; at $N=5$ it is $2.8$ times $c_{5,1}$.  Over
$\alpha_s(m_\tau)\in[0.30,0.33]$ the CIPT tolerances change by only $-8\%$ to
$+20\%$.  For the present validation question, FOPT is therefore the
stronger test.

This comparison should not be confused with the all-order FOPT--CIPT
separation.  In modern Borel analyses~\cite{Hoang:2020mkw,Hoang:2021nlz,
Gracia:2023qdy}, one may define an asymptotic separation $\Delta_W^{\rm AS}$
as the difference between specified all-order Borel sums associated with the
two perturbative organizations.  Its value is prescription- and
model-dependent, and it receives information from the whole Borel singularity
structure~\cite{Benitez-Rathgeb:2022yqb,Benitez-Rathgeb:2022hfj,
Golterman:2023oml}.  Determining $c_{5,1}$ improves the finite-order input but
does not set $\Delta_W^{\rm AS}$ to zero and does not decide which
organization best approximates a selected Borel sum.

\section{Direct Shanks/Pad\'e forecasts and balance}
\label{sec:pade}

\subsection{The local geometric model}

We truncate the Adler series~\eqref{eq:adler-series} at the last known order and
write its partial sums as
\begin{equation}
 \Dhat_N(a)=\sum_{n=1}^{N}t_n,\qquad t_n=c_{n,1}a^n .
 \label{eq:adler-partial-sums}
\end{equation}
The first Shanks transformation, equivalently Aitken's $\Delta^2$
process~\cite{Weniger:1989}, acts on the last three partial sums as
\begin{equation}
 e_1(\Dhat_{N-1})
 =\frac{\Dhat_N\Dhat_{N-2}-\Dhat_{N-1}^2}{\Dhat_N-2\Dhat_{N-1}+\Dhat_{N-2}}
 =\Dhat_N+\frac{t_N^2}{t_{N-1}-t_N}
 =\Dhat_N+\sum_{j\geq1}t_N\,\varrho_N^{\,j},
 \label{eq:aitken}
\end{equation}
where $\varrho_N\equiv t_N/t_{N-1}=a\,c_{N,1}/c_{N-1,1}$.
It is exact when the remainder is a single geometric mode, and it
completes the series with a geometric tail whose term ratio is the last
observed one.  Expanding in $a$,
$e_1(\Dhat_{N-1})=\Dhat_N+(c_{N,1}^2/c_{N-1,1})\,a^{N+1}+\cO(a^{N+2})$, so the
first unknown coefficient is forecast by extending the last observed
coefficient ratio by one order:
\begin{equation}
 \frac{\chat_{N+1,1}^{\rm dir}}{c_{N,1}}=\frac{c_{N,1}}{c_{N-1,1}}
 \qquad\Longrightarrow\qquad
 \chat_{N+1,1}^{\rm dir}=\frac{c_{N,1}^2}{c_{N-1,1}},
 \qquad N\geq2,\quad c_{N-1,1}\neq0 .
 \label{eq:direct-shanks}
\end{equation}
The coupling cancels, as in the ratio fits of Section~\ref{sec:bj-values}.

The same formula follows from matching a one-pole rational function.  Writing
$e_1(\Dhat_{N-1})=\Dhat_{N-2}+t_{N-1}^2/(t_{N-1}-t_N)$ and clearing the
denominator $c_{N-1,1}-c_{N,1}a$ shows that $e_1(\Dhat_{N-1})/a$ is the
$[N-2/1]$ Pad\'e approximant of $\Dhat/a$ in the convention of
Eq.~\eqref{eq:pade-definition} below: it reproduces all $N$ known coefficients,
in accordance with the counting rule $L+M=N-1$ of
Eq.~\eqref{eq:pade-counting}.  Its denominator $1-(c_{N,1}/c_{N-1,1})\,a$
generates $\chat_{k+1,1}=(c_{N,1}/c_{N-1,1})\,\chat_{k,1}$ for $k\geq N$, whose
first step is Eq.~\eqref{eq:direct-shanks}.  This is why the estimate is
called a direct Pad\'e/Shanks forecast: the first Shanks element and the
one-pole $[N-2/1]$ approximant encode the same local geometric assumption.
For $N=4$ this is the $[2/1]$ approximant, one of the balanced choices of
Eq.~\eqref{eq:balanced-condition}, and
\begin{equation}
 \chat_{5,1}^{\rm dir}=\frac{c_{4,1}^2}{c_{3,1}}=378.03.
 \label{eq:direct-c5}
\end{equation}

The same algebra can instead be applied to each FOPT moment
series~\eqref{eq:triangular-map}, with $c_{n,1}\to d_n[W]$ and $a\to a_0$,
giving the moment-space forecast $\dhat_{N+1}^{\rm dir}[W]=d_N[W]^2/d_{N-1}[W]$.
Converted into an Adler forecast through Eq.~\eqref{eq:gap}, it is in general
weight-dependent, because the ratio extrapolation does not commute with the
linear map.  For $\Wtau$ at $N=4$,
$\dhat_5^{\rm dir}[\Wtau]=127.079^2/26.366=612.50$ and, by
Eq.~\eqref{eq:d5-affine}, $\chat_{5,1}=612.50-307.783=304.71$.  This is the
constant-ratio counterpart of the BJ linear-ratio estimate
$283$ in Eq.~\eqref{eq:bj-c5}.  Both are moment-space forecasts and should not
be conflated with the common-Adler value~\eqref{eq:direct-c5}; the same holds
for the Levin-transform estimate $c_{5,1}=278^{+27}_{-19}$ of
Ref.~\cite{Abbas:2025dpm}, which transforms the FOPT series of $\delta^{(0)}$.
The separate four-parameter BJ Borel-model variant, fitted to
$c_{1,1},\ldots,c_{4,1}$ only, forecasts $280.12$
(Section~\ref{sec:cvrc}).  These numbers test
different maps and are not contradictory.

The direct forecast~\eqref{eq:direct-c5} is not a theorem about QCD.  Renormalon-dominated coefficients grow
factorially~\cite{Beneke:1998ui}, approximately as
\begin{equation}
 c_{n,1}\sim K A^n\Gamma(n+b),
 \label{eq:factorial-growth}
\end{equation}
so their ratio is not constant but grows linearly with $n$.  The direct
forecast can still be a useful local diagnostic, but its geometric remainder
model is too rigid to represent several competing infrared and ultraviolet
renormalon modes.

We can divide coefficients by the factorial growth before applying the same
idea.  In the conventional Borel variable~\cite{Beneke:1998ui,Beneke:2008ad} for which the leading infrared
renormalons occur at positive integers, define
\begin{equation}
 B[\Dhat](u)=\sum_{n=0}^{\infty}
 \frac{c_{n+1,1}}{\beta_0^n n!}\,u^n,
 \qquad
 -Q^2\frac{\dd a}{\dd Q^2}=\beta_0a^2+\cO(a^3),
 \label{eq:borel-transform}
\end{equation}
where $\beta_0=9/4$ for $n_f=3$ and $a=\alpha_s/\pi$.  If
$b_n=c_{n+1,1}/(\beta_0^n n!)$, direct continuation gives
\begin{equation}
 \widehat b_4=\frac{b_3^2}{b_2}
 \qquad\Longrightarrow\qquad
 \chat_{5,1}^{\rm Borel\text{-}dir}
 =\frac{4}{3}\frac{c_{4,1}^2}{c_{3,1}}=504.04.
 \label{eq:borel-direct-c5}
\end{equation}
The large shift between $378.03$ and $504.04$ is evidence that four
coefficients do not identify a unique asymptotic representation.

\subsection{Balanced and unbalanced Pad\'e approximants}

Pad\'e approximants have been applied to the $\tau$ series in
Ref.~\cite{Boito:2018rwt}.  Let $F(z)=f_1z+\cdots+f_Nz^N+\cO(z^{N+1})$ be
any of the series above, with $N$ known coefficients.  After factoring the
known zero at
$z=0$, approximate $F(z)/z$ by
\begin{equation}
 [L/M](z)=\frac{p_0+p_1z+\cdots+p_Lz^L}
 {1+q_1z+\cdots+q_Mz^M}.
 \label{eq:pade-definition}
\end{equation}
There are $L+1+M$ unknown parameters.  Matching the $N$ known coefficients of
$F(z)/z$ therefore requires
\begin{equation}
 L+M=N-1.
 \label{eq:pade-counting}
\end{equation}
To obtain a rational continuation with at least one denominator pole, one also
requires $M\geq1$.  A common near-diagonal selection rule is
\begin{equation}
 L+M=N-1,\qquad M\geq1,\qquad |L-M|\leq1.
 \label{eq:balanced-condition}
\end{equation}
The first two conditions follow from parameter counting and the desire to
continue beyond a polynomial.  The last is a design criterion, not a theorem:
it prevents either numerator or denominator from carrying almost all degrees
of freedom.  When $N=4$, the balanced choices are $[2/1]$ and $[1/2]$.

A $[3/0]$ approximant is the input cubic itself.  Because it has no denominator
and no higher Taylor coefficients, it supplies no nontrivial next-order
forecast.  A $[0/3]$ approximant can forecast, but it is strongly unbalanced:
its constant numerator forces all nontrivial structure into the denominator,
making the result especially sensitive to low-order noise and spurious poles.
Unbalanced approximants are not mathematically forbidden.  They are valuable
as stability diagnostics, but should not silently be assigned the same status
as a near-diagonal family.

\section{The defect-sign obstruction}
\label{sec:obstruction}

We define the one-step direct-forecast defect
\begin{equation}
 \Delta_n=c_{n,1}-\frac{c_{n-1,1}^2}{c_{n-2,1}},
 \qquad c_{0,1}\equiv1.
 \label{eq:defect-definition}
\end{equation}
It measures how far the new coefficient lies from the direct
forecast~\eqref{eq:direct-shanks} built from the preceding two; the convention
$c_{0,1}\equiv1$ defines $\Delta_2$.  The known Adler coefficients give
\begin{equation}
 (\Delta_2,\Delta_3,\Delta_4)=(0.640,\,3.682,\,24.323).
 \label{eq:known-defects}
\end{equation}
All three are positive and their magnitudes increase.  A direct Pad\'e/Shanks
forecast sets the next defect to zero.  More generally, call a map
defect-sign-preserving if it returns $\widehat\Delta_{N+1}\geq0$ when the
supplied defects are positive and increasing.  Such a map must satisfy
\begin{equation}
 \chat_{5,1}\geq\frac{c_{4,1}^2}{c_{3,1}}=378.03.
 \label{eq:defect-bound}
\end{equation}

The three-weight validation at the previous order does not reveal this
restriction.  At $N=4$, the ten-percent tolerance in
Table~\ref{tab:fopt-tolerance} accepts
\begin{equation}
 \chat_{4,1}\in[46.925,51.227],
 \label{eq:c4-window}
\end{equation}
which is equivalent to
\begin{equation}
 \widehat\Delta_4\in[22.172,26.474].
 \label{eq:defect-window}
\end{equation}
The entire accepted interval has positive defect.  Hence the validation test
can accept a sign-preserving continuation even though that continuation is
structurally unable to reach a value around $280$ at the next order.  A value
$c_{5,1}=280$, for example, corresponds to
\begin{equation}
 \Delta_5=280-378.03\simeq-98.03,
 \label{eq:required-reversal}
\end{equation}
a sign reversal not encoded by the local geometric pattern.

This gives a sharp dichotomy:
\begin{quote}
Either $c_{5,1}\geq378.03$, or every defect-sign-preserving continuation of
the common Adler sequence misses $c_{5,1}$, including those that pass the
ten-percent withheld-order test at $N=4$.
\end{quote}
The statement is conditional and falsifiable.  It does not prove that
$c_{5,1}$ is large.  It proves that obtaining a smaller number requires a map
outside the sign-preserving class, hence additional declared structure.

\begin{figure}[H]
\centering
\includegraphics[width=0.88\textwidth]{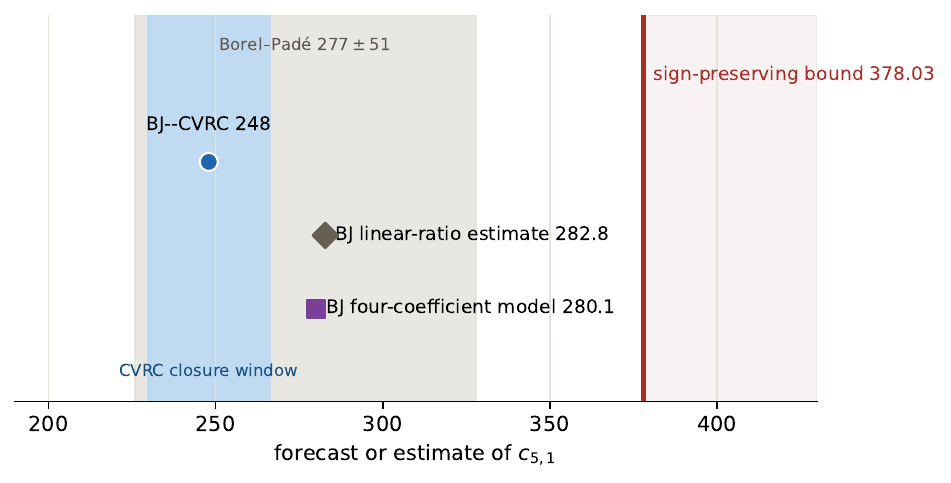}
\caption{Compact numerical comparison.  The blue circle is the BJ--CVRC
central value $247.98\simeq248$; the light-blue interval is its structural
closure window $[229.8,266.2]$, not the full adopted uncertainty $248\pm80$.
The diamond is the BJ moment-space linear-ratio estimate
$282.823\simeq283$, which is an input to the main five-parameter BJ model,
whereas the square is the separate four-coefficient BJ-model forecast
$280.12$.  The gray band is the independent Borel--Pad\'e range
$277\pm51$, and the red line is the direct common-Adler sign-preserving
boundary $c_{4,1}^2/c_{3,1}=378.03$.}
\label{fig:partition}
\end{figure}

Figure~\ref{fig:partition} gives the compact comparison needed here: it
separates the three reference values near the phenomenological range from the
direct common-Adler sign-preserving boundary.  The complete method-by-method
classification is deferred to the companion manuscript.

An ensemble average requires special care.  An average over Pad\'e
approximants, Borel normalizations, or renormalization schemes need not inherit
a nonlinear inequality obeyed by each member.  It can lie where no individual
map lies.  Therefore the Borel--Pad\'e estimate $277\pm51$ of
Ref.~\cite{Boito:2018rwt} is not in conflict with
Eq.~\eqref{eq:defect-bound}; it is an ensemble result, not one
defect-sign-preserving transformation of the four common-Adler inputs.

\section{Renormalons, OPE ambiguities, and what resummation cannot remove}
\label{sec:renormalons}

Suppose the Borel transform of the perturbative Adler function, equivalently
the perturbative Wilson coefficient $C_0$ in the OPE, has a positive-axis
singularity
\begin{equation}
 B[\Dhat](u)\underset{u\to p}{=}
 \frac{N_p}{(p-u)^{1+\nu}}
 \left[1+\cO(p-u)\right],
 \qquad p>0.
 \label{eq:renormalon-singularity}
\end{equation}
Expanding the leading term about $u=0$ gives
\begin{equation}
 \frac{1}{(p-u)^{1+\nu}}
 =\frac{1}{p^{1+\nu}}
 \sum_{n=0}^{\infty}
 \frac{\Gamma(n+1+\nu)}{\Gamma(1+\nu)n!}
 \left(\frac{u}{p}\right)^n.
 \label{eq:branch-expansion}
\end{equation}
With the Borel convention of Eq.~\eqref{eq:borel-transform}, this produces the
factorial large-order behavior
\begin{equation}
 c_{n+1,1}\sim
 \frac{N_p}{\Gamma(1+\nu)}
 \frac{\beta_0^n\Gamma(n+1+\nu)}{p^{n+1+\nu}}.
 \label{eq:renormalon-asymptotics}
\end{equation}
The position $p$ controls the exponential rate, the exponent $\nu$ controls a
power of $n$, and $N_p$ is the residue or normalization of that renormalon
mode.

In the same convention, the formal Borel--Laplace reconstruction is
\begin{equation}
 \Dhat_{\mathcal P}(a)=\frac{1}{\beta_0}
 \int_{\mathcal C_{\mathcal P}}\dd u\,
 e^{-u/(\beta_0a)}B[\Dhat](u),
 \label{eq:borel-laplace}
\end{equation}
where $\mathcal C_{\mathcal P}$ specifies how the positive-axis singularities
are bypassed.  Expanding Eq.~\eqref{eq:borel-transform} inside the integral
immediately reproduces $\sum_{n\geq0}c_{n+1,1}a^{n+1}$, which checks all
normalization factors.
Because the singularity~\eqref{eq:renormalon-singularity} lies on the
integration path,
one must specify whether the contour passes above or below the cut.  The two
lateral sums differ by an imaginary ambiguity of the parametric form
\begin{equation}
 \Dhat_{+}-\Dhat_{-}\propto
 i\,e^{-p/(\beta_0a)}a^{-\nu}.
 \label{eq:lateral-ambiguity}
\end{equation}
Renormalization-group evolution converts the exponential into a power
correction of order $(\Lambda^2/Q^2)^p$, associated with OPE dimension
$D=2p$~\cite{Beneke:1998ui}.  The OPE~\cite{Wilson:1969zs} separates the
observable schematically as
\begin{equation}
 \mathcal R(Q)=C_0(Q;\mathcal P)+
 \sum_D\frac{C_D(Q;\mathcal P)}{Q^D}
 \langle O_D(\mathcal P)\rangle,
 \label{eq:OPE-prescription}
\end{equation}
where $\mathcal P$ denotes the common scheme and Borel prescription.  The
ambiguity of the lateral sum of $C_0$ is cancelled by the corresponding
ambiguity in the operator matrix element (and, where appropriate, its Wilson
coefficient).  The physical observable is prescription independent; the
separate perturbative and condensate pieces are not.  Renormalon-free
definitions of the gluon condensate~\cite{Benitez-Rathgeb:2022yqb,
Benitez-Rathgeb:2022hfj} and gradient-flow formulations of the
OPE~\cite{Beneke:2023wkq,Beneke:2025hlg} make this compensation explicit.

Pad\'e approximation and sequence transformations can approximate analytic
continuation, reveal candidate singularities, and improve numerical estimates
within a chosen prescription.  They cannot eliminate the physical need for a
prescription, because no manipulation of the perturbative coefficients alone
defines the compensating nonperturbative matrix element.  A quoted
principal-value Borel sum is a consistent and useful convention, but it is a
convention.  This is why the BJ--CVRC uncertainty below must not be interpreted
as removing OPE prescription dependence.

\section{A Beneke--Jamin-type ansatz and cross-validated closure}
\label{sec:cvrc}

The defect obstruction identifies what a successful continuation near the
usual range must contain: a mechanism capable of reversing the defect sign.
Ultraviolet renormalons on the negative Borel axis generate sign-alternating
contributions and provide such a mechanism.  The natural precedent is the
renormalon model of BJ~\cite{Beneke:2008ad}; related Adler-function models
are discussed in Refs.~\cite{Caprini:2009vf,DescotesGenon:2010cr,Jamin:2010cv,
Caprini:2010tf,Caprini:2011ya,Beneke:2012vb}.

The construction used here is not the full
BJ model.  It retains the leading singularity locations and
renormalization-group exponents, replaces full cuts by pure powers, and omits
some subleading cut coefficients and Wilson-coefficient information.  We call
it a \emph{Beneke--Jamin-type ansatz}.  ``CVRC'' names the
cross-validated-renormalon-closure protocol applied to that ansatz; it does not
name a new physical Borel model.  The physical construction is denoted
\emph{CVRC-minimal}: at each order it contains only the finite set of declared
singularity and polynomial columns needed to match the available coefficients.
A separate \emph{CVRC-complete}
construction, containing the full singularity tower available in the exactly
solvable large-$\beta_0$ limit, is used only as an all-order validation
laboratory and is not the model used for the physical forecast.

The algebra of closure can be seen without committing to every basis function.
We write the model as
\begin{equation}
 B_{\rm ansatz}(u;\boldsymbol\theta,\rho)
 =\sum_{j=1}^{K}\theta_j\phi_j(u)+\rho\,\phi_{\rm cl}(u),
 \label{eq:closure-model}
\end{equation}
where $\phi_j$ are fixed renormalon or polynomial basis functions,
$\boldsymbol\theta$ are amplitudes determined by known coefficients, and
$\rho$ is one deliberately unclosed amplitude.  Matching $K$ known Borel
coefficients gives a linear system
\begin{equation}
 A\boldsymbol\theta+\rho\,\boldsymbol v=\boldsymbol b,
 \qquad
 \boldsymbol\theta(\rho)=A^{-1}\boldsymbol b-\rho A^{-1}\boldsymbol v.
 \label{eq:closure-linear-system}
\end{equation}
Any next coefficient extracted linearly from the model is therefore affine in
$\rho$,
\begin{equation}
 \chat_{K+1,1}(\rho)=\kappa_{0,K}+\kappa_{1,K}\,\rho,
 \label{eq:affine-forecast}
\end{equation}
with intercept $\kappa_{0,K}$ and slope $\kappa_{1,K}$ fixed by the declared
structure and the $K$ known coefficients.

Because perturbative coefficients form an ordered sequence, the available
fold is a one-step rolling-origin calibration.  The lower-order family built
from $c_{1,1},\ldots,c_{N-1,1}$ withholds the next known coefficient and fixes
$\rho_\star$ through
\begin{equation}
 \widehat c_{N,1}(\rho_\star)=c_{N,1}.
 \label{eq:ordered-calibration}
\end{equation}
The calibration target is therefore reproduced exactly after $\rho_\star$ is
fixed and is not independent validation.  The nontrivial test is whether the
same closure amplitude is compatible, within the declared tolerance, with the
lower-order fold and all three moment weights.  Equivalently, the admissible
$\rho$ intervals obtained at the known orders must have a nonempty
intersection; an empty intersection rejects the declared singularity
structure.  At the physical order the calibrated central value coincides
algebraically with the exact four-parameter fit of the pure-power ansatz and
contains no information beyond those four inputs.  CVRC adds the
acceptance-or-rejection test, the admissible interval, external benchmarks,
and a calibrated method uncertainty; it does not turn the fit into a
determination.  We use ``cross-validated'' only in this restricted cross-order
consistency sense, not in the statistical sense of an independent validation
sample.

For the declared structure the admissible set is nonempty,
$\rho\in[-0.0679,-0.0529]$ around $\rho_\star=-0.0604$, and the surviving
pure-power BJ-type construction gives
\begin{equation}
 c_{5,1}^{\rm BJ\text{-}CVRC}
 =248\pm30\;({\rm structure})\pm74\;({\rm method})
 =248\pm80.
 \label{eq:cvrc-result}
\end{equation}
The two components have different meanings.  The structural component
combines linearly the admissible closure window, which maps to
$c_{5,1}\in[229.8,266.2]$, and a variation of the unknown $u=3$ exponent
between $3.00$ and $3.58$, which moves the central value over $[235.3,257.8]$.
This gives $^{+28}_{-31}$, symmetrized to $\pm30$.  The method component is the
error of the transformation itself, and it dominates.  The genuinely blind
version of the protocol, with $\rho$ fixed one order lower from $c_{3,1}$,
postdicts $\chat_{4,1}=34.43$, which is $29.9\%$ below the known value.  In the
large-$\beta_0$ laboratory the same minimal tower is first defined at $N=5$,
where its one-order-ahead error is $+23.7\%$; blind tests on the four-loop
pole-mass coefficient~\cite{Marquard:2015qpa} give $2.3$--$11.9\%$, but that
series is dominated by the nearby $u=1/2$ renormalon and is an optimistic
bound.  We therefore set the method component to $30\%$ of the central value,
$74$, the larger of the $23.7\%$ error at the first defined minimal-tower
large-$\beta_0$ order and the $29.9\%$ magnitude of the strictly recursive
physical miss.  This choice brackets the only same-observable blind failure
instead of assuming that its relative size improves at the next order.
Because the laboratory errors are positive while the physical blind miss is
negative, no sign-informed correction is justified and the band is symmetric.
The quoted $80$ is the quadrature combination
$\sqrt{30^2+74^2}=79.8$, rounded to the nearest ten.

We observe that the central value is compatible with the Borel--Pad\'e estimate
$277\pm51$~\cite{Boito:2018rwt}, with the Levin-transform estimate
$278^{+27}_{-19}$~\cite{Abbas:2025dpm}, and with the four-coefficient forecast
$280.12$ of the four-parameter BJ predictive variant~\cite{Beneke:2008ad},
which lies $0.40$ times the total uncertainty above it.  Agreement does not,
however, establish uniqueness.  The $32$-unit difference from that variant is
entirely subleading
structure: replacing pure powers by BJ's cut corrections adds $9.6$, and the
gluon-condensate Wilson coefficient a further $22.6$.  With that structure
declared, however, the admissible set is empty (the closure drifts by $0.236$
between folds, against $0.051$ for pure powers), and the strictly recursive
postdiction is $\chat_{4,1}=244.5$, five times the true value.  This is a
useful warning: adding theoretically correct asymptotic information can worsen
low-order prediction if the asymptotic regime has not begun.

An empty admissible set is also what makes CVRC a test of the declared
structure.  In a misspecification scan the leading infrared singularity is
moved from $u=2$ to positions where the Adler function has none, and the
leading ultraviolet singularity from $u=-1$ to $u=-3/2$ or $u=-2$.  A leading
singularity at $u=1$ is rejected, as it should be: the massless Adler function
has no $u=1$ renormalon, because no gauge-invariant operator of dimension two
enters its OPE.  Both ultraviolet displacements and $u=5/2$ are also rejected.
The test is not perfectly discriminating: a leading singularity at $u=3/2$
survives, with $c_{5,1}=290.4$, and the ultraviolet exponent is resolved only
coarsely.

The large-$\beta_0$ limit~\cite{Beneke:1998ui,LovettTurner:1994hx,Ball:1995ni}
used above retains and reorganizes a subset of fermion-bubble insertions so
that renormalon singularities are explicit.  It is a controlled laboratory for
large-order behavior, not a quantitatively exact replacement for full QCD.  In
it the complete singularity tower and the minimal tower reach stable accuracy
at different orders.  The complete tower meets ten percent persistently from
$N=6$, whereas the minimal tower used for the physical construction does so
only from $N=8$; keeping this distinction prevents a successful asymptotic
test from being overinterpreted at physical order four.  The companion
paper~\cite{CompanionPaper} documents all of these tests.

\section{Experimental relevance and future facilities}
\label{sec:future}

The coefficient $c_{5,1}$ is not an observable: it multiplies $a^5$ in the
massless Adler function and, in a specified renormalization scheme, can be
fixed unambiguously only by a genuine six-loop calculation.  Future $\tau$
data can nevertheless test its
phenomenological consequences.  Belle~II and the proposed Super Tau--Charm
Facility (STCF) can improve exclusive branching fractions, inclusive spectral
distributions, strange--nonstrange channel separation, detector unfolding,
and bin-to-bin covariance matrices~\cite{Kou:2018nap,STCF:2023CDR}.  Belle
data already enter the recent $\alpha_s$ analysis of
Ref.~\cite{Boito:2025pwg}.  More precise spectra reduce moment uncertainties
and may permit additional weights or tighter validation criteria.  They do
not, however, change the algebraic resolution in
Eq.~\eqref{eq:fifth-order-resolution} when the weights and tolerance
$\epsilon$ are held fixed.

The FCC-ee $Z$-pole programme would provide $\mathcal{O}(10^{11})$
$Z\to\tau^+\tau^-$ decays, together with precise high-scale QCD
observables~\cite{Dam:2021FCCtau,FCC:2025FSR}.  If sufficiently inclusive
hadronic $\tau$ spectra and their covariance matrices can be reconstructed,
they would furnish complementary low-scale moment tests; high-scale
observables would instead test the running and global consistency of
$\alpha_s$.  Neither programme directly measures $c_{5,1}$.

A controlled forecast of $c_{5,1}$ is therefore useful for estimating the
first omitted FOPT contribution, assigning correlated truncation uncertainties
consistently across different weights, and preventing perturbative model
dependence from being mistaken for experimental tension.  

\section{Scope, limitations, and guide to the companion paper}
\label{sec:long-paper-guide}

Each conclusion includes its hypotheses.  The transfer identity is exact for
one common Adler forecast mapped to
normalized FOPT moments with the lower coefficients held fixed; independent
moment-space forecasts give weight-dependent Adler values
(Section~\ref{sec:pade}).  The defect-sign obstruction constrains individual
sign-preserving continuations of the common Adler sequence; it does not apply
to moment-space estimates such as the BJ linear-ratio value $283$, and it need not be
inherited by an ensemble average.  The CIPT criterion is also one-dimensional,
but it lacks the unit diagonal and acts on the contour-improved partial sum,
the closest analogue of the coefficient-level FOPT test.  Finally,
$c_{5,1}=248\pm80$ is conditional on the declared pure-power ansatz and the
closure rule; the band is nonstatistical and does not remove the OPE
prescription dependence (Section~\ref{sec:renormalons}).

This paper is self-contained for the physics, exact identities, and numerical
conclusion.  The companion paper~\cite{CompanionPaper} provides the full
renormalization-group recursion and FOPT map matrices; the direct,
Borel-normalized, Pad\'e, Wynn, Levin, and Brezinski~$\theta$
transformations~\cite{Levin:1973,Smith:1979,Smith:1982,Weniger:1989},
including algebraic coincidences among nominally different methods; the
singularity-structure and misspecification scans with their closure
intervals; the large-$\beta_0$ and pole-mass benchmarks; and the numerical
contour integration for CIPT.  The accompanying reproducibility package
regenerates the letter's numerical audit and figure with two independent
implementations.

\section{Conclusions}

Several spectral weights are physically well motivated, since they differ in
pinching, OPE projection, and renormalon sensitivity.  A common
Adler-coefficient forecast nevertheless carries a single unknown error.
Because the FOPT Adler-to-moment map is triangular with unit diagonal, that
error is transferred unchanged to every normalized moment, and a simultaneous
relative test reduces to one tolerance set by the smallest moment
coefficient.  For the three weights used here, a ten-percent fifth-order test
resolves $c_{5,1}$ only to about $\pm24$; the closest CIPT analogue, which acts
on the partial sum, is weaker by more than an order of magnitude.

The known coefficients also expose a structural obstruction.  The defects of
the direct Pad\'e/Shanks continuation are positive and increasing, so every
sign-preserving continuation of the Adler sequence gives $c_{5,1}\geq378$,
whereas the usual range near $280$ requires the next defect to be negative.
Reaching that range requires a mechanism for the reversal.  A simplified
BJ-type renormalon ansatz supplies one, and CVRC rejects the ansatz
if no single closure reproduces the two highest known orders.  The resulting
conditional estimate,
$c_{5,1}=248\pm30\,(\mathrm{struct})\pm74\,(\mathrm{method})=248\pm80$,
is broad by design.  Its calibrated central value is the exact
four-parameter pure-power fit; the evidential content lies in the
admissibility tests and in the uncertainty assignment.  It is compatible with
the usual range, but four coefficients do not determine the fifth precisely.

No Pad\'e approximation, sequence transformation, or renormalon fit removes
the OPE prescription dependence, because the perturbative Borel ambiguity
cancels only against the matrix-element ambiguity in a common scheme and
prescription.  The decisive step would be a direct calculation of $c_{5,1}$.
Until then, declared structural assumptions, falsifiable validation, and
honest uncertainty bands are more informative than precision generated by an
underconstrained model.

\section*{Acknowledgments}
G.A.\ is grateful to Antonio Pich for crucial discussions during his visit to
the Instituto de F\'isica Corpuscular (IFIC), Universitat de Val\`encia--CSIC,
which shaped the core of this work.  He also thanks IFIC for its warm
hospitality and stimulating research environment.

\end{document}